\documentclass[prl,aps,twocolumn]{revtex4-1}

\usepackage[utf8]{inputenc}

\usepackage{amsmath,amssymb,latexsym,epsfig,graphics,epsf}
\usepackage{color} 
\usepackage{hyperref}

\newcommand{\be}{\begin{equation}}
\newcommand{\ee}{\end{equation}}

\begin{document} 
	\renewcommand{\i}{\operatorname{i}} 

	\title{Solid-liquid coexistence of the noble elements. I. Theory illustrated by the case of argon}
	\author{Aditya N. Singh}
	\affiliation{Theoretical Chemistry Institute and Department of Chemistry, University of Wisconsin-Madison, 1101 University Avenue, Madison, Wisconsin 53703, USA}
	\author{Jeppe C. Dyre}
	\author{Ulf R. Pedersen} 
	\email{ulf@urp.dk}
	\affiliation{{\it Glass and Time}, IMFUFA, Department of Science and Environment, Roskilde University, P. O. Box 260, DK-4000 Roskilde, Denmark} 
	
	\date{\today}
	
	\begin{abstract} 
		The noble elements constitute the simplest group of atoms. At low temperatures or high pressures they freeze into the face-centered cubic (fcc) crystal structure (except helium). We perform molecular dynamics using the recently proposed simplified {\it ab initio}
		atomic (SAAP) potential [Deiters and Sadus, J. Chem. Phys. 150, 134504 (2019)] . This potential is parameterized using data from accurate {\it ab initio} quantum mechanical calculations by the coupled-cluster approach on the CCSD(T) level. We compute the fcc freezing lines for Argon and find a great agreement with the experimental values. At low pressures, this agreement is further enhanced by using many-body corrections. Hidden scale invariance of the potential energy function is validated by computing lines of constant excess entropy (configurational adiabats) and shows that mean square displacement and the static structure factor are invariant. These lines (isomorphs) can be generated from simulations at a single state-point by having knowledge of the pair potential. The isomorph theory for the solid-liquid transition is used to accurately predict the shape of the freezing line in the pressure-temperature plane, the shape in the density-temperature plane, the entropy of melting and the Lindemann parameters along the melting line. We finally predict that the body-centered cubic (bcc) crystal is stable at high pressures.
	\end{abstract}
	
	\maketitle 
	
	\section{Introduction}
	
	Thermodynamic and transport properties of condensed matter systems at a given temperature and density are determined by their potential energy functions. For a class of systems, the potential energy function exhibits a hidden scale invariance that makes the phase diagram effectively one dimensional, thus density and temperature collapses into a single parameter. In this paper we investigate argon (Ar) using a potential proposed recently from accurate {\it ab initio} calculations. We conclude that the energy surface obeys hidden scale invariance (in the investigated part of the phase diagram), and show that this fact can be used to predict the shape of the melting lines. In the companion paper (II) we apply the theory derived here to the other noble elements Ne, Kr and Xe.
	
	\section{Realistic potential energy surface}
\begin{figure}
		\begin{center}
			\includegraphics[width=0.49\textwidth]{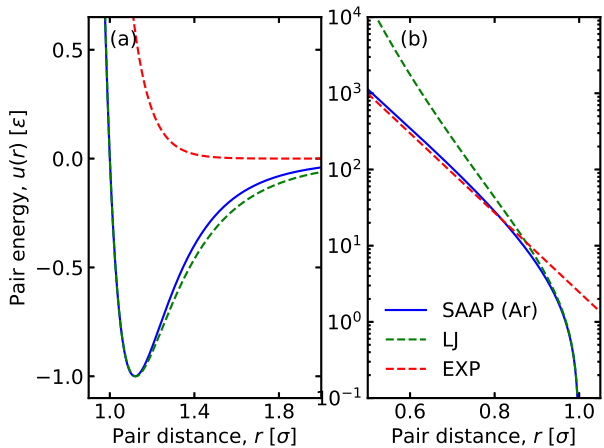}
		\end{center}
		\caption{\label{fig:SAAP} (a) The SAAP pair potential for Argon (red; solid), and LJ (green; dashed) and the EXP (red; dashed) pair potentials. (b) The SAAP pair potential on a logarithmic scale.
		}
	\end{figure}

	We investigate the simplified {\it ab initio}
	atomic (SAAP) potential recently suggested by Deiters and Sadus \cite{deiters2019}. This potential is parameterized for the noble elements Ne, Ar, Kr and Xe from quantum mechanical calculations using the coupled cluster approach \cite{konrad2010,bartlett2007} on the CCSD(T) theoretical level \cite{nasrabad2004}. This approach has been referred to as the ``gold standard'' of quantum chemistry \cite{carsky2010} and is shown to give accurate prediction for the noble elements \cite{bartlett2007,lopez2004}.
	Below we consider monatomic systems of $N$ particles of mass $m$ confined to a volume $V$ with periodic boundaries (a three-dimensional torus) with the number density $\rho=N/V$. Let ${\bf R} = ( {\bf r}_1, {\bf r}_2, {\bf r}_3,\ldots,{\bf r}_N)$ be the collective coordinate vector. The potential energy surface is defined as a sum of pair potentials
	\begin{equation}\label{Eq:U}
	U({\bf R}) = \sum_{i>j}^N\varepsilon v(|{\bf r}_i-{\bf r}_j|/\sigma)
	\end{equation}
	where the SAAP pair potential is
	\begin{equation}
	v(r) = \frac{a_0\exp(a_1r)/r+a_2\exp(a_3r)+a_4}{1+a_5r^6}.
	\end{equation}
	The parameters for Ar are $\varepsilon/k_B=$143.4899372 K, $\sigma=0.3355134529$ \AA, $a_0=65214.64725$, $a_1=-9.452343340$, $a_2=-19.42488828$, $a_3=-1.958381959$, $a_4=-2.379111084$, $a_5=1.051490962$ \cite{deiters2019}. These coefficients are determined by fitting to results of the above mentioned {\it ab initio} calculations on dimers \cite{konrad2010}. The pair potential is truncated and shifted at $r_c=4$ in units of $\sigma$.
	An advance of the SAAP potential is that it is computationally efficient while accurately representing the underlying {\it ab initio} calculations \cite{deiters2019}.
	Figure \ref{fig:SAAP}(a) shows the SAAP pair potentials of Ar in units of $\varepsilon$. The potential has been parameterized to have the same minimum as the Lennard-Jones (LJ) potential shown as a green dashed line: $4[r^{-12}-r^{-6}]$. The LJ potential is too steep at short distances (Fig.\ \ref{fig:SAAP}(b)) as noted by Thiel and Alder \cite{thiel1966}. The red dashed line is the exponential repulsive (EXP) pair potential $4\cdot10^5\exp(-12r)$. The SAAP potential is approximated by the EXP potential \cite{bacher2014,dyre2016,bacher2018,pedersen2019} as short distances, see Fig. \ref{fig:SAAP}(b). This is consistent with the interpretation of high-pressure compression experiments (shock Hugoniots) [references].
	
	Simulations were conducted using the RUMD software package \cite{rumd}. We studied systems of $N=5120$ particles in an elongated orthorhombic simulation cell where the box length in the $y$ and $z$ directions are identical, and the box length in the $x$ direction is 2$\frac{1}{2}$ times longer. We perform molecular dynamics for $2^{22}\simeq4\times10^6$ steps after equalization using a leap-frog time-step of $0.004\sigma\sqrt{m/\varepsilon}$. This results in a simulation time of about $1.7\times10^4\sigma\sqrt{m/\varepsilon}=33$ ns. The temperature $T$ and/or pressure $p$ is keep constant using the Langevin type dynamics suggested by Grønbech-Jensen et al.\ \cite{gronbech2014}. 
	
	\begin{figure}
		\begin{center}
			\includegraphics[width=0.49\textwidth]{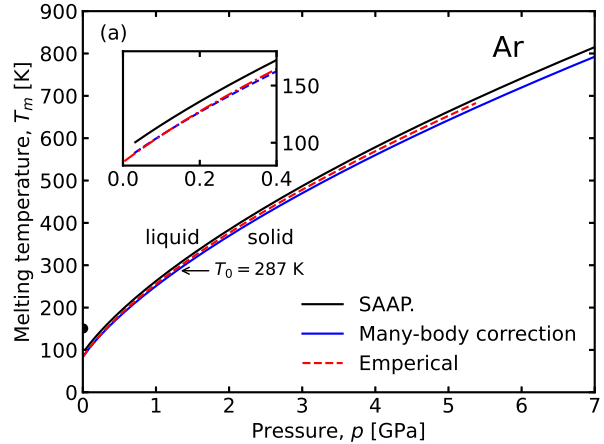}
			\includegraphics[width=0.49\textwidth]{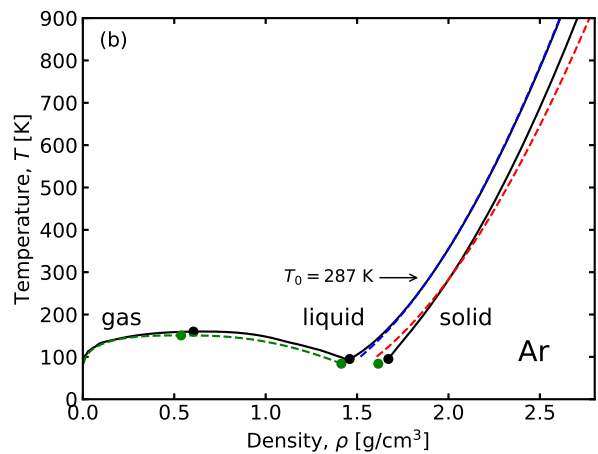}
			\caption{\label{fig:argonPhases} (a) The solid-liquid coexistence line of Argon in the $pT$-plane. The black solid line is the melting temperature $T_m(p)$ computed for the SAAP potential by first using the interface pinning method at $T_0=287$ K, and then use a fourth-order Runge-Kutta integration of the Clausius-Clapeyron identity to determine remaining coexistence points. The solid blue line is a mean-field correction to missing many-body interactions of the SAAP potential (see main text), and the red dashed line is the empirical melting line by Datchi et.\ al \cite{datchi2000}. The black dot is the gas-liquid critical point. The inset is a zoom-in on low pressures. (b) The phase diagram in the $\rho T$-plane. The black solid lines are the boundaries of the phases. The three black dots are the critical point, and the triple point for the liquid and solid, respectively. The green dots and dashed lines correspond to empirical values. The blue and red dashed lines are liquid and a solid isomorphs based on the SAAP potential, respectively.}
		\end{center}
	\end{figure}

	The coexistence lines between liquid and the fcc solid are determined as follows: First, we use the interface-pinning method \cite{pedersen2013} to compute the solid-liquid chemical potential difference $\Delta\mu$ at temperature $T_0 = 2\varepsilon/k_B$ (287 K) for a range of fcc lattice constants corresponding to different pressures. The $\Delta \mu$ in the interface-pinning method computed from the thermodynamic force on a solid-liquid interface in a simulation with an auxiliary potential that biases the system towards two-phase configurations. Two-phase simulations were done eight times longer than bulk simulations to account for slow fluctuations of the solid-liquid interface. We determine that the coexistence pressure ($\Delta\mu=0$) is $p=22.591(4)\varepsilon/\sigma^3$ at $T_0 = 2\varepsilon/k_B$. The number in parenthesis indicates the estimated error on the last digit. Other coexistence points are then determined by numerical integration along temperatures on the coexistence line using the fourth-order Runge-Kutta (RK4) algorithm \cite{numerical_recipes} where the required slopes $dp/dT$ are computed from isobaric simulations of a solid and a liquid using the Clausius-Clapeyron relation: if $\Delta V_m=V_\textrm{liquid}-V_\textrm{solid}$ is the volume difference between liquid and solid, $\Delta S_m=(U_\textrm{liquid}-U_\textrm{solid}+p\Delta V_m)/T$ is the entropy difference between the two phases, then the local slope is computed as $dP/dT=\Delta S_m/\Delta V_m$. This recipe for computing solid-liquid coexistence lines was suggested in Reference \cite{pedersen2019}. As a consistency check we confirm that the gradient of the central difference of the computed melting line agrees with $\Delta S_m/\Delta V_m$ as computed above (see Supplementary material). The result is shown as a solid black line on Fig.\ \ref{fig:SAAP}(a) together with the empirical values \cite{datchi2000} shown with a red dashed line. The agreement is good; however, as seen in the inset, the SAAP gives a slight overestimate of the coexistence temperature at a given pressure. This is likely due to missing many-body interactions of the SAAP potential. To investigate this, we apply a mean-field correction that only depends on the average density of a bulk phase (following the suggestion by Deiters and Sadus given in Reference \cite{deiters2019b}). For a correction of the coexistence pressure we take the different densities of the phases into account. Let $\bar\rho=2/(v_s+v_l)$ be the average density between the two phases at a given coexistence point ($T^\text{SAAP},p^\text{SAAP}$) computed with the SAAP potential. The corrected state-point is then ($T_c,p_c$)=($\varepsilon_cT^\text{SAAP},\varepsilon_cp^\text{SAAP}$) where $\varepsilon_c=1-\frac{\lambda\nu\bar\rho}{\varepsilon\sigma^6}$ is an energetic correction parameter of the $\varepsilon$ in Eq.\ \ref{Eq:U}, $\nu/\varepsilon\sigma^9=0.0687536$ is the Axilrod-Teller-Muto parameter \cite{axilrod1943} given in Reference \cite{deiters2019b}, and $\lambda$ is a fudge parameter that we set to unity for simplicity. The solid blue line on Fig.\ \ref{fig:argonPhases} shows the corrected melting line. The correction is small, but explains the deviations from the experimental melting line at low pressures (inset on Fig.\ \ref{fig:argonPhases}(a))). At high pressures, however, the uncorrected melting line is better than the corrected. This suggests that many-body interactions are less important at high pressures. For the remainder of the paper we will ignore many-body corrections, but expect that the inclusion of the many-body effects will give rise to some quantitative changes to our conclusions. We leave such investigations to future studies.
	
	\section{Hidden scale invariance}
\begin{figure}
		\begin{center}
			\includegraphics[width=0.49\textwidth]{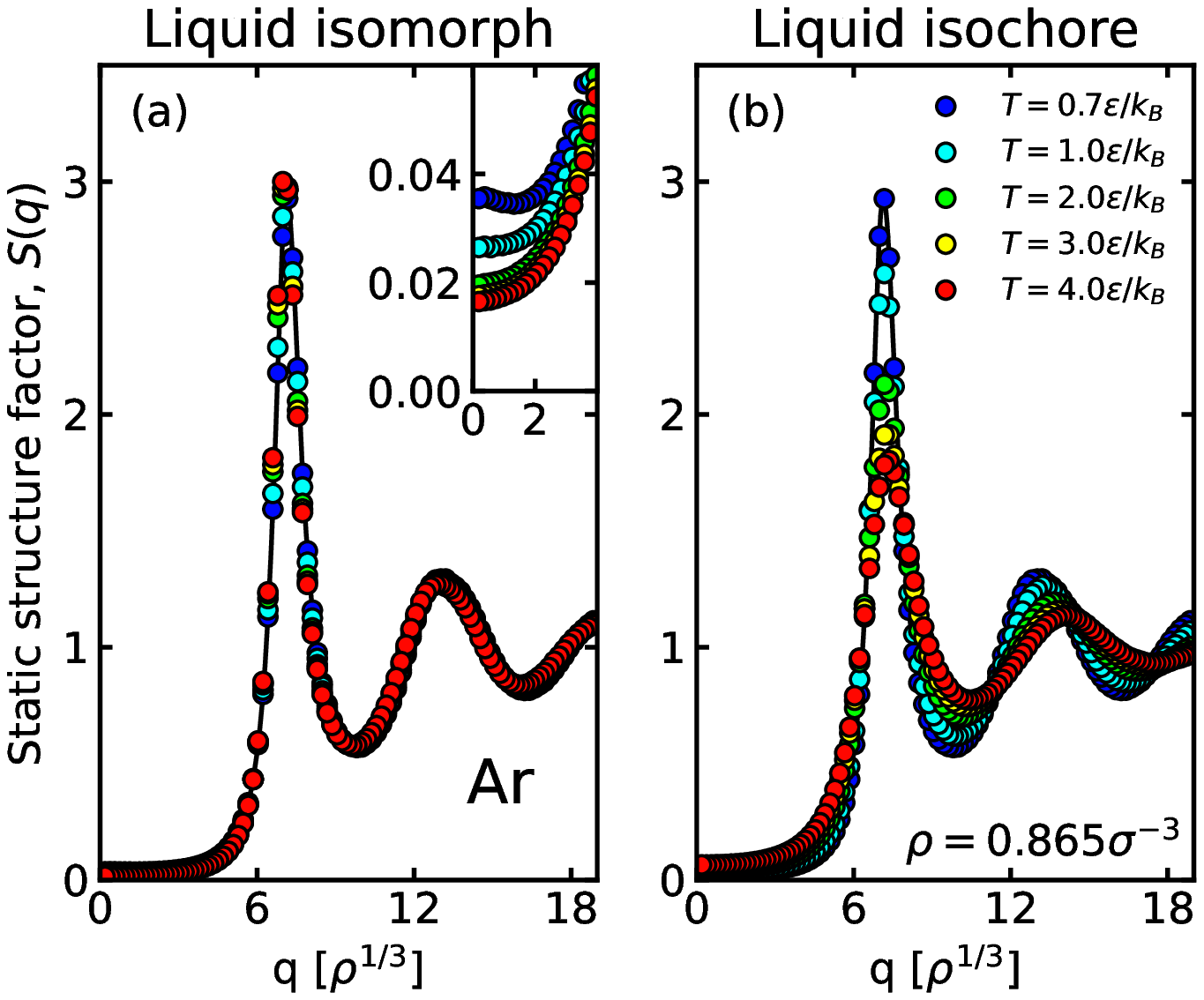}
			\includegraphics[width=0.49\textwidth]{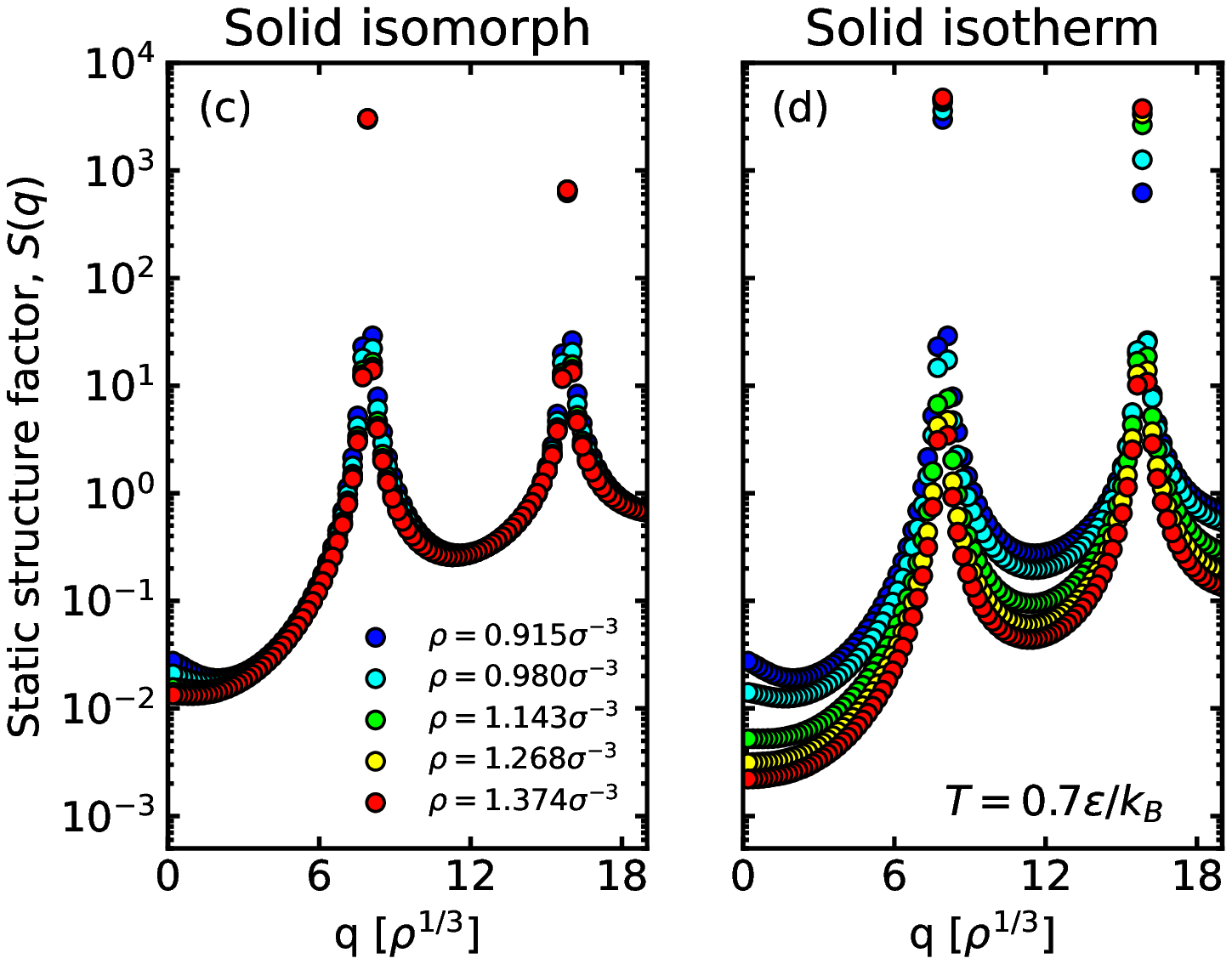}
		\end{center}
		\caption{\label{fig:Sq} (a) The static structure factor, $S({\bf q})$, along the liquid isomorph of SAAP Ar near the freezing line (blue dashed line on Fig.\ \ref{fig:argonPhases}(b)). The first axis shows $q=|{\bf q}|$ in units of $\rho^{1/3}$. The structure is invariant along the isomorph. The solid line is a guide to the eye. The inset zoom-in on $S({\bf q})$ for short ${\bf q}$ vectors demonstrating that the isothermal compressibility is not isomorph invariant because $S(\bf{0})$ is not. (b) $S({\bf q})$ along the liquid isochore with the same density as the state point on the isomorph with temperature $T=0.7 \varepsilon/k_B$. (c) $S({\bf q})$ with the wave-vector ${\bf q}$ pointing in the elongated [100] direction of the fcc solid along isomorphic state-points (red dashed line on Fig.\ \ref{fig:argonPhases}(b)). The first two Bragg-peaks are shown. (d) $S({\bf q})$ of the fcc solid along the $T = 0.7\varepsilon/k_B$ isotherm for the same range of densities as the corresponding isomorph.}
	\end{figure}

\begin{figure}
		\begin{center}
			\includegraphics[width=0.49\textwidth]{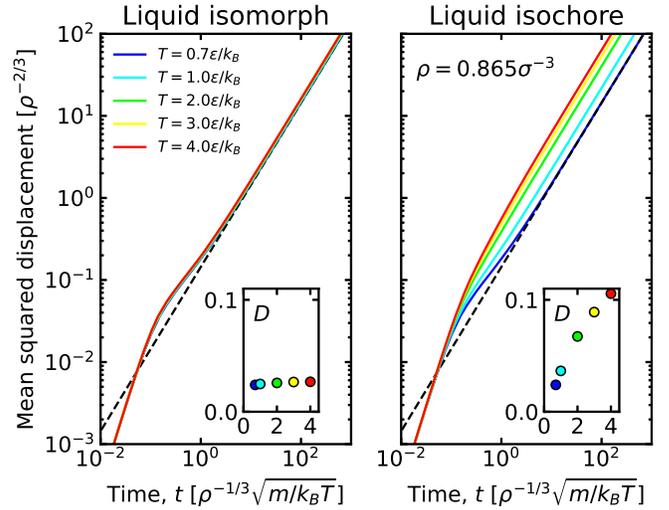}
		\end{center}
		\caption{\label{fig:msd} (a) Mean squared displacement along the liquid SAAP isomorph of Ar near the freezing line (blue dashed line on Fig.\ \ref{fig:argonPhases}(b)). The collapse of the data demonstrate that dynamics are invariant along the isomorph. The inset shows isomorph invariance of the diffusion constant in reduced units, $\sqrt{k_BT/m}\rho^{-1/3}$. The diffusion constant is computed from the long time limit of the mean squared displacement. (b) Mean squared displacement along the liquid isochore with the same density as the state point on the isomorph with temperature $T=0.7 \varepsilon/k_B$.}
	\end{figure}

	The following gives a brief introduction to the theory of systems with hidden scale invariance, known as the isomorph theory \cite{gnan2009,schroeder2014,dyre2014,dyre2016}, and applies it to the Ar parametrization of the SAAP potential. Consider two configurations ${\mathbf R}_a$ and ${\mathbf R}_b$ where $U({\bf R}_a)<U({\bf R}_b)$. If the energy surface has hidden scale invariance for those configurations, it follows that $U(\lambda {\bf R}_a)<U(\lambda  {\bf R}_b)$ where $\lambda$ determines the magnitude of an affine scaling of the particle positions and thus the density. From this definition of hidden scale invariance, it follows that there are lines in the phase diagram, referred to as isomorphs, where structure, dynamics, and some thermodynamics quantities are invariant in units that are reduced by a combination of particle mass $m$, the number density $\rho$ and the kinetic energy $k_BT$ \cite{schroeder2014}. The isomorphs are defined as lines where the excess entropy $S_\textrm{ex}$ is constant, i.e. a configurational adiabat. Here, ``ex'' refer to the entropy {\it in excess} of the ideal gas entropy: $S_\textrm{ex}=S-S_\textrm{id}$. This scaling with excess entropy was first suggested by Rosendeld in 1977 \cite{rosenfeld1977}, but have recently gained renewed interest \cite{krekelberg2007,chakraborty2007,mausbach2018,dyre2018,bell2019}.
	A configurational adiabat is only referred to as an isomorph for state-points with hidden scale-invariance and thus invariant structure (iso-morf is the greek word for same-shape).
	The slope of a configurational adiabat (and an isomorph) in the double logarithmic temperature-density plane,
	\begin{equation}\label{eq:gamma}
	\gamma \equiv \left.\frac{\partial \ln T}{\partial \ln \rho}\right|_{S_\textrm{ex}},
	\end{equation}
	can be computed from the fluctuations of virial and potential energy in the constant $NVT$ ensemble as
	$
	\gamma = \langle\Delta W\Delta U\rangle/\langle(\Delta U)^2\rangle
	$ \cite{gnan2009}.
	Here $\langle \ldots\rangle$ is the thermodynamic average in the constant $NVT$ ensemble and $\Delta$ denotes the deviation from the mean. The dashed lines on Fig.\ \ref{fig:argonPhases}(b) show a liquid and a solid isomorph computed by numerical integration of Eq.\ \ref{eq:gamma} using the fourth-order Runge-Kutta method from a reference state point $(T_0,\rho_0)$ \cite{attia2020}. The isomorphic state points can also be found by the direct isomorph check (DIC) method. This method relies on the fact that the structure is invariant along an isomorph, allowing the isomorph to be computed from configurations at the reference state point. Figure \ref{fig:Sq}(a) shows that the structure is indeed invariant by investigating the static structure factor $S({\bf q})=\langle |\rho_{\bf q}|^2 \rangle$ where $\rho_{\bf q}=\sum_n^N\exp(\mathrm{i}{\bf q}\cdot{\bf r}_n)/\sqrt{N}$. For comparison, Fig.\ \ref{fig:Sq}(b) shows $S({\bf q})$ for state-points along an isochore starting near the triple point. Figs.\ \ref{fig:Sq}(c) and \ref{fig:Sq}(d) show $S({\bf q})$ for the fcc solid along state-points of the isomorph and a isotherm, respectively. As for the liquid, the structure is invariant along the isomorph. We note that the long-wavelength (short ${\bf q}$-vector) limit of the structure factor does not scale well (inset on Fig.\ \ref{fig:Sq}(a) and Fig.\ \ref{fig:Sq}(c)). This limit is proportional to the isothermal compressibility, which is not an isomorph invariant \cite{heyes2019}. Figure \ref{fig:msd}(a) shows that dynamics is invariant along the liquid isomorph by investigating the mean squared displacement and the diffusion constant (inset) computed by the long-time limit (dashed line). Figure \ref{fig:msd}(b) shows the same along an isochore.
	
\begin{figure}
		\begin{center}
			\includegraphics[width=0.49\textwidth]{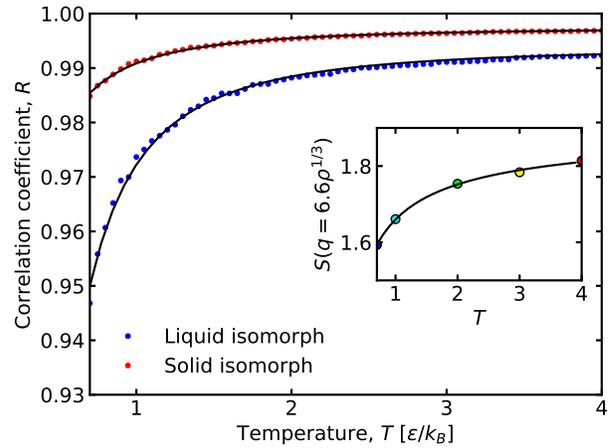}
		\end{center}
		\caption{\label{fig:correlation_coefficient} The Pearson correlation coefficient $R$ between virial $W$ and potential energy $U$ along the liquid isomorph (blue) and the solid isomorph (red) of SAAP Ar. The solid line is a guide to the eye. The inset shows the static structure factor $S(|q=6.6\rho^{1/3})$ as a function of temperature. The correlation coefficient approaches unity with increasing temperature. The insert shows the static structure factor at $q=6.6\rho^{1/3}$ showing that the structure becomes more invariant when the correlation coefficient is low. The solid lines are guides to the eye.}
	\end{figure}

	In an $NVT$ simulation the virial is given by $W({\bf R})=-\sum_{i>j}\varepsilon|{\bf r}_i-{\bf r}_j|u^{(1)}(|{\bf r}_i-{\bf r}_j|/\sigma)/3\sigma$ where $u^{(1)}$ is the first derivative of the pair potential with respect to the reduced pair distance. The virial is referred to as the ``potential part of the pressure'' since the pressure $p$ is given by the relation $pV=Nk_BT+ \langle W\rangle$.
	Systems with hidden scale-invariance are sometimes referred to as ``strongly correlating'' \cite{pedersen2008} since the fluctuations of virial and potential energy are strongly correlated in the $NVT$ ensemble. Figure \ref{fig:correlation_coefficient} show the Pearson correlation coefficient $R$ between $W({\bf R})$ and $U({\bf R})$ for the liquid (blue points) and the crystal isomorph (red points). The correlation is strong as expected from the invariant structure and dynamics (Figs.\ \ref{fig:Sq} and \ref{fig:msd}). $R>0.94$ for all investigated state-points. The correlation increases with increasing temperature (and density). This is consistent with the fact that the structure is more invariant on the high-temperature part of the configurational adiabat (inset on Fig.\ \ref{fig:correlation_coefficient}).

\begin{figure}
		\begin{center}
			\includegraphics[width=0.49\textwidth]{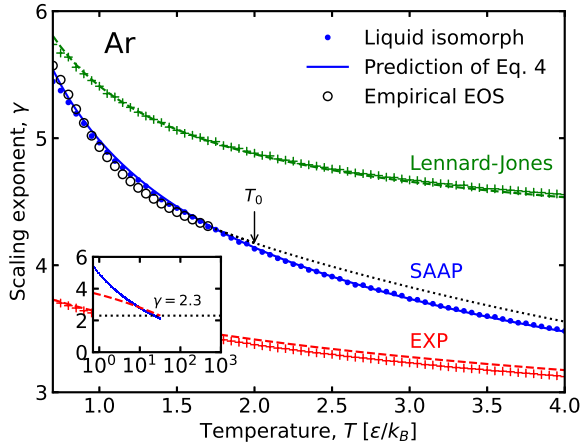}	
		\end{center}
		\caption{\label{fig:gamma} The blue dots show the scaling exponent of SAAP Ar along the liquid isomorph. The solid blue line is an estimate that only refer to the pair-potential: Eq.\ \ref{eq:pair_gamma} with $\Lambda(S_\textrm{ex})=1.047$. The $\Lambda$ value is chosen to give the correct $\gamma$ at the reference temperature $T_0=2\varepsilon/k_B$ (indicated with an arrow). The agreement is good for all the investigated state points. The green dashed line is the predictions of the LJ potential, and the green $+$ symbols are the exponents computed in simulations at the state-points of the liquid isomorph of the LJ potential. The red dashed line and $+$'s is the same for the EXP potential. The open circles are values from the empirical Ar EOS by Tegeler et al.\ \cite{tegeler1999}. The inset show $\gamma$ along the liquid isomorph at high temperatures. At $T=21\varepsilon/k_B=3000$ K the exponent goes below $\gamma<2.3$, suggesting that the stable crystal is bcc at high temperatures.
		}
	\end{figure}

	If the pair potential follows an inverse power-law ($r^{-n}$) then the isomorphs are given by $\rho^\gamma/T=$const.\ where $\gamma=n/3$ is constant \cite{gnan2009}. Thus $\gamma$ is referred to as the ``density scaling exponent''. In general, however, $\gamma$ is state point dependent \cite{sanz2019}. It has been demonstrated that for many systems with pair interaction, including the LJ and the EXP systems, that the exponent can be approximated by fitting an {\it effective} inverse power-law to the pair potential as some distance \cite{bohling2014}. For the dense phases (liquid and solid) this results in the expression
	\begin{equation}\label{eq:pair_gamma}
	\gamma(\rho,S_\textrm{ex}) \simeq -\frac{2}{3}-\left.\frac{r}{3}\frac{u^{(3)}(r)}{u^{(2)}(r)}\right|_{r=\Lambda(S_\textrm{ex})\rho^{-1/3}/\sigma}
	\end{equation}
	where $u^{(i)}(r)$ is the $i$th derivative of the pair potential with respect to $r$ and $\Lambda(S_\textrm{ex})$ is a free parameter for a given isomorph expected to be close to unity. Under the assumption that $\Lambda$ is the same for nearby isomorphs, $\gamma$ is only a function of density to a good approximation. Figure \ref{fig:gamma} shows the true $\gamma$ of Eq.\ \ref{eq:gamma} (dots) and the $\gamma$ estimated from the pair interactions (solid lines). The agreement is excellent. Thus, any isomorph can be computed from a single reference state-point since it can be computed by integrating Eq.\ \ref{eq:pair_gamma} with $\Lambda$ determined at the state point by calculating $\gamma$ from the fluctuations.

	In this paper we investigate the fcc solid, however, it well-known that they can form other structures such as hexagonal closed packing (hcp) at higher pressures than what we investigate in this paper \cite{ferreira2008}.
	At high pressures (density and temperature) the exponent $\gamma$ approaches that of the EXP potential and becomes smaller. As such, the pair potential becomes long-ranged compared to the interparticle distance. When $\gamma<2.3$ it is expected that the body centered cubic (bcc) crystal becomes thermodynamically stable compared to the closed packed crystals (fcc and hcp) \cite{hoover1972,khrapak2012,hummel2015}. Hoover and coworkers \cite{hoover1972} explained the fcc-bcc-fluid triple point for many metals by the fact that the effective pair potential becomes soft. It was recently shown that the EXP pair potential has a fcc-bcc-fluid triple point \cite{pedersen2019} located where $\gamma=2.33(3)$ \cite{bacher2020}. Since the potential of the noble elements are approximated by the EXP potential at high densities, we expect that the noble elements have an fcc-bcc-fluid triple point where $\gamma\simeq2.3$. For Argon we estimate the triple point fcc-bcc-fluid triple point to be found at $T=21\varepsilon/k_B=3000$ K (inset on Fig. \ref{fig:gamma}). Belonoshko and coworkers \cite{belonoshko2001,belonoshko2008} have argued that such a triple point exists for Xe based on theoretical calculations and re-interpretations of experiments. The experiments presented in Reference \cite{ross2005} were, however, unsuccessful in detecting such a triple point.
	We note that $\gamma$ for the LJ potential is 4 in the high-pressure limit, thus, the fcc-bcc-fluid triple points is never reach \cite{bondarev2011}. The LJ potential, however, does not describe the noble elements at high pressures since it is too harsh. The EXP high-pressure limit of the pair interactions also suggest a reentrence temperature above which no crystalline phase is stable (see $T^\star$ in Ref.\ \cite{bacher2020}).
	
	The density-scaling exponent can be determined from thermodynamic data as the ratio between the excess pressure coefficient $\beta_V^\textrm{ex}\equiv (\partial W/\partial T)_V/V=\kappa^\textrm{ex}_T\alpha^\textrm{ex}_T$ and the excess isochoric heat capacity per volume $c_V^\textrm{ex}\equiv(\partial U/\partial T)_V/V$: $\gamma = \beta_V^\textrm{ex}/c_V^\textrm{ex}$ \cite{gnan2009}. These are usually not directly available from experiments, but using standard thermodynamic relations this can be re-written as $\gamma=[\gamma_G-k_B/c_V]/[1-3k_B/2c_V]$ where $\gamma_G = \left.\frac{\partial \ln T}{\partial \ln \rho}\right|_{S} = \alpha_pK_T/c_V$ is the well-studied thermodynamic Grüneisen parameter \cite{gruneisen1912,mausbach2014,mausbach2016,nagayama2011}. Within the accuracy of the Dulong-Petit approximation, $c_V\simeq3k_B$, then $\gamma \simeq 2\gamma_G-2/3$. The value of the Grüneisen parameter is $\gamma_G=2.9$ \cite{amoros1988,mausbach2018} near the gas-liquid-solid triple point of Ar. This corresponds to $\gamma=5.1$, and is in good agreement with the value obtained by the SAAP potential (Fig.\ \ref{eq:gamma}). Amoros et.\ al \cite{amoros1988} notice that $\gamma_G$ is only a function of density to a good approximation. This is explained by the fact that $\Lambda$ in Eq.\ \ref{eq:pair_gamma} is close to unity for all $S_\textrm{ex}$, making $\gamma$ as well as $\gamma_G$ functions of density exclusively. Thus, $\gamma_G$ is also only a function of density. This is only expected to be true in dense phases, i.e. the liquid and the solid. In the ideal gas limit only temperature is expected to be releveant as illustrated for the EXP potential in Reference \cite{EXPI}. The reason for this is that the typical collision distance of gas particles only depends on temperature.
	In Fig.\ \ref{fig:gamma} (open circles) we compare the $\gamma$ along the liquid isomorph of the SAAP potential to that of the the empirical equation of state (EOS) by Tegeler et al.\ \cite{tegeler1999}. (This EOS is implemented into the CoolProp \cite{bell2014} software library by Bell et.\ al.) The agreement is good.

	\section{Theory of the melting line}
	
	\begin{figure}
		\begin{center}
			\includegraphics[width=0.49\textwidth]{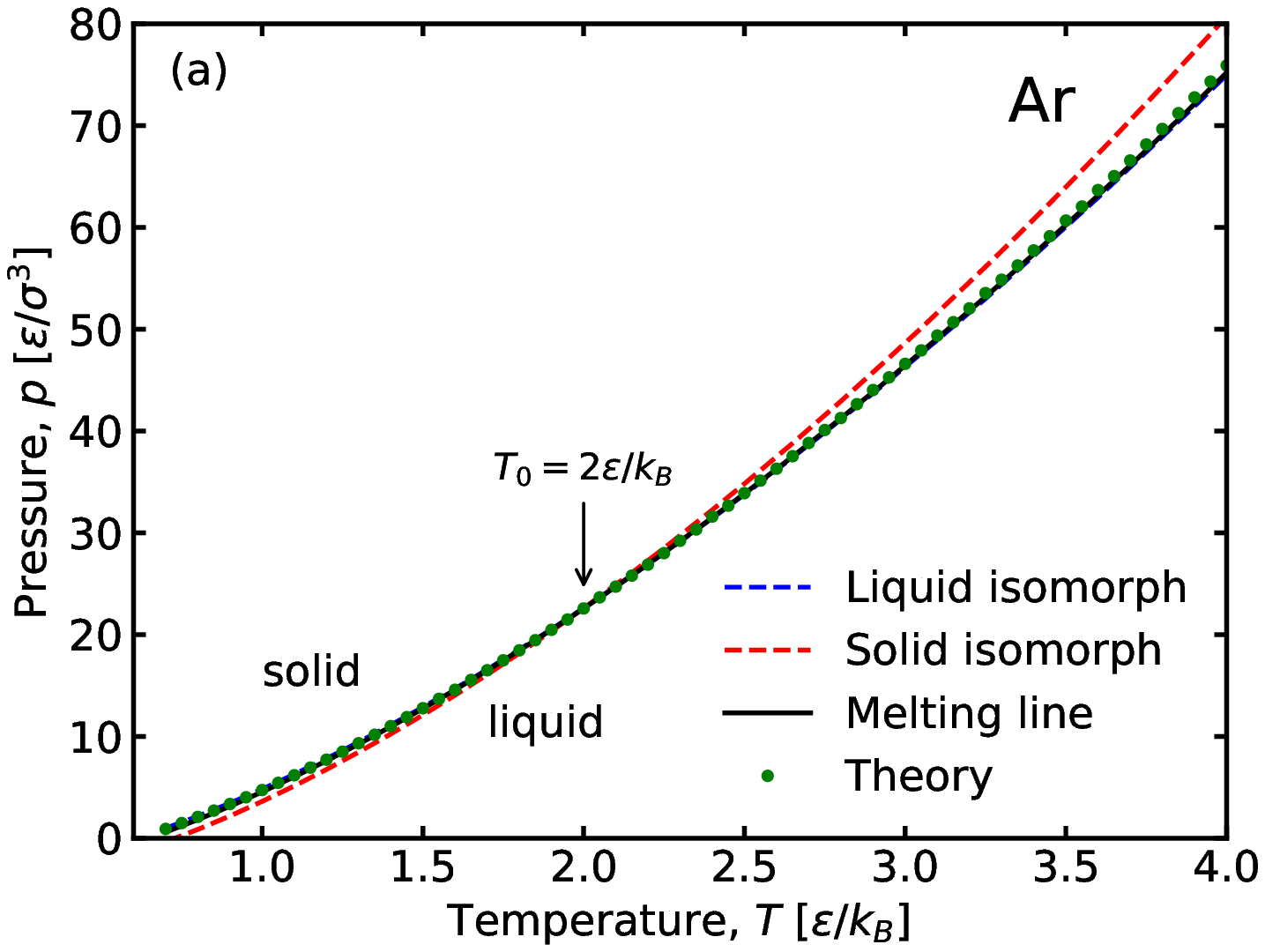}
			\includegraphics[width=0.49\textwidth]{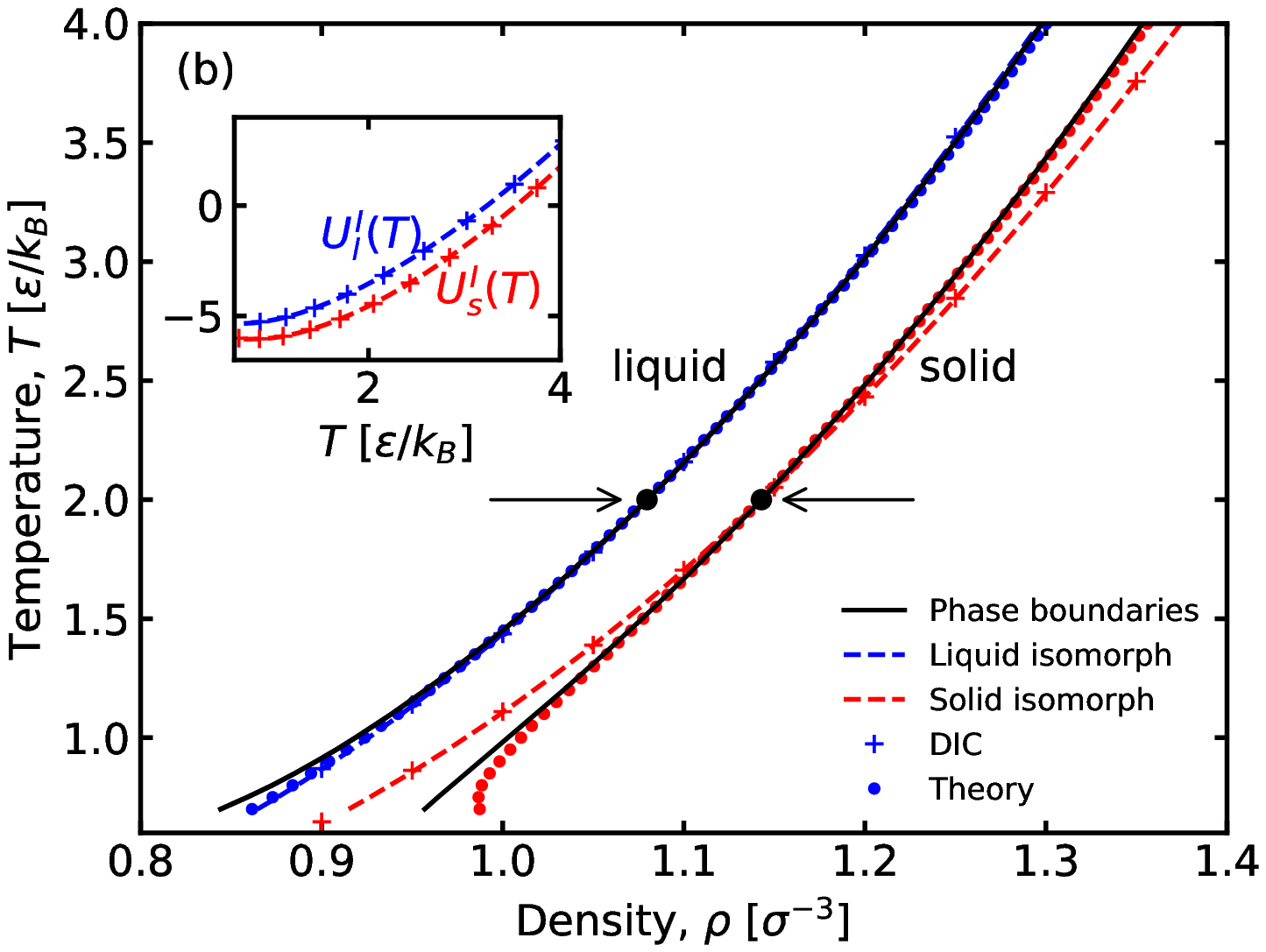}		
		\end{center}
		\caption{\label{fig:melting_theory} Applying the melting theory presented in Reference \cite{pedersen2016}. The prediction is made at the reference state point at temperature $T_0=2\varepsilon/k_B$ indicated with arrows. The insert show the potential energy per particle along the liquid (blue) and solid (red) isomorph. The $+$'s on both the main figure and the inset is results of the DIC method.}
	\end{figure}
	
	\begin{figure}
		\begin{center}
			\includegraphics[width=0.49\textwidth]{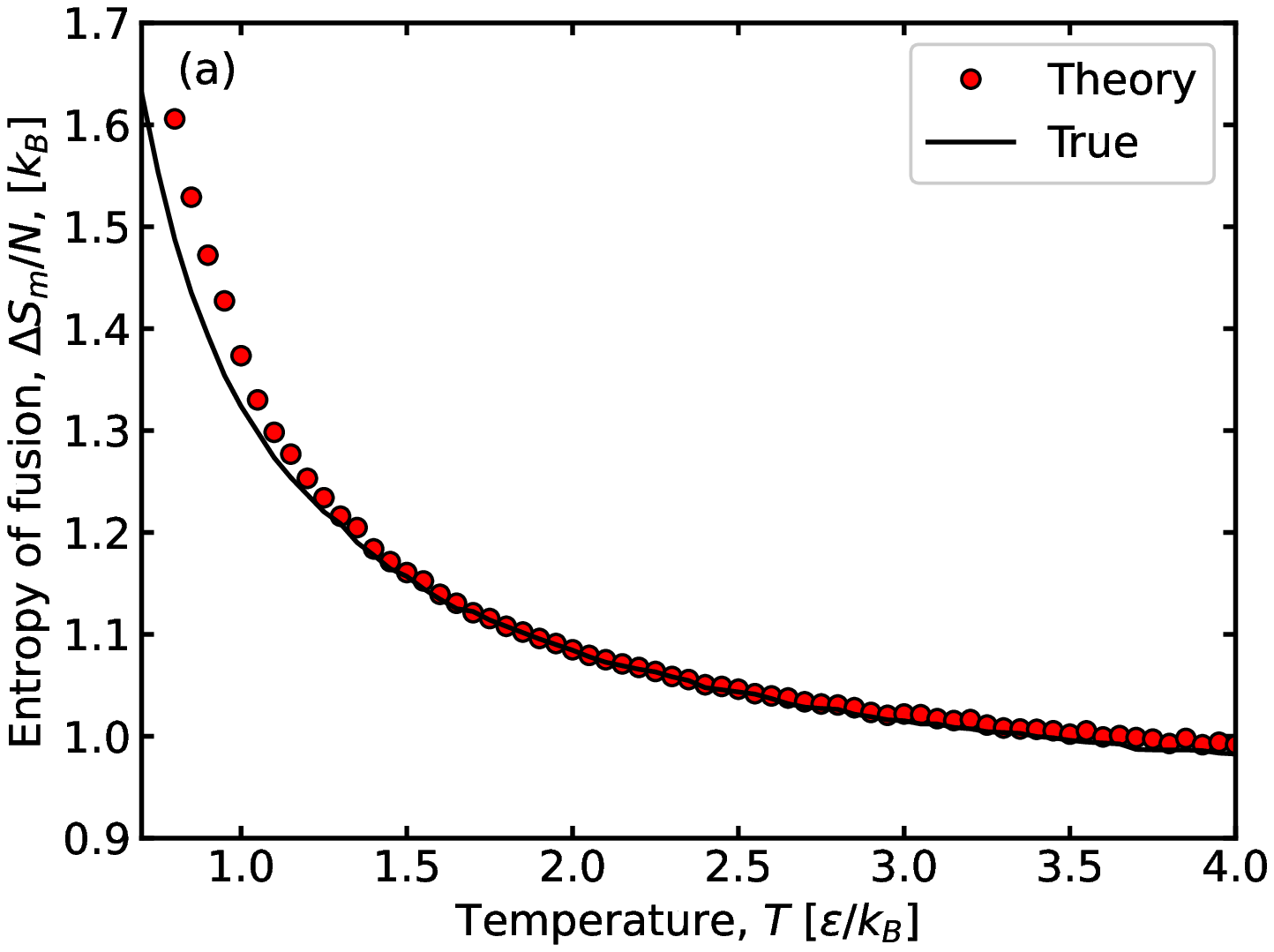}
			\includegraphics[width=0.49\textwidth]{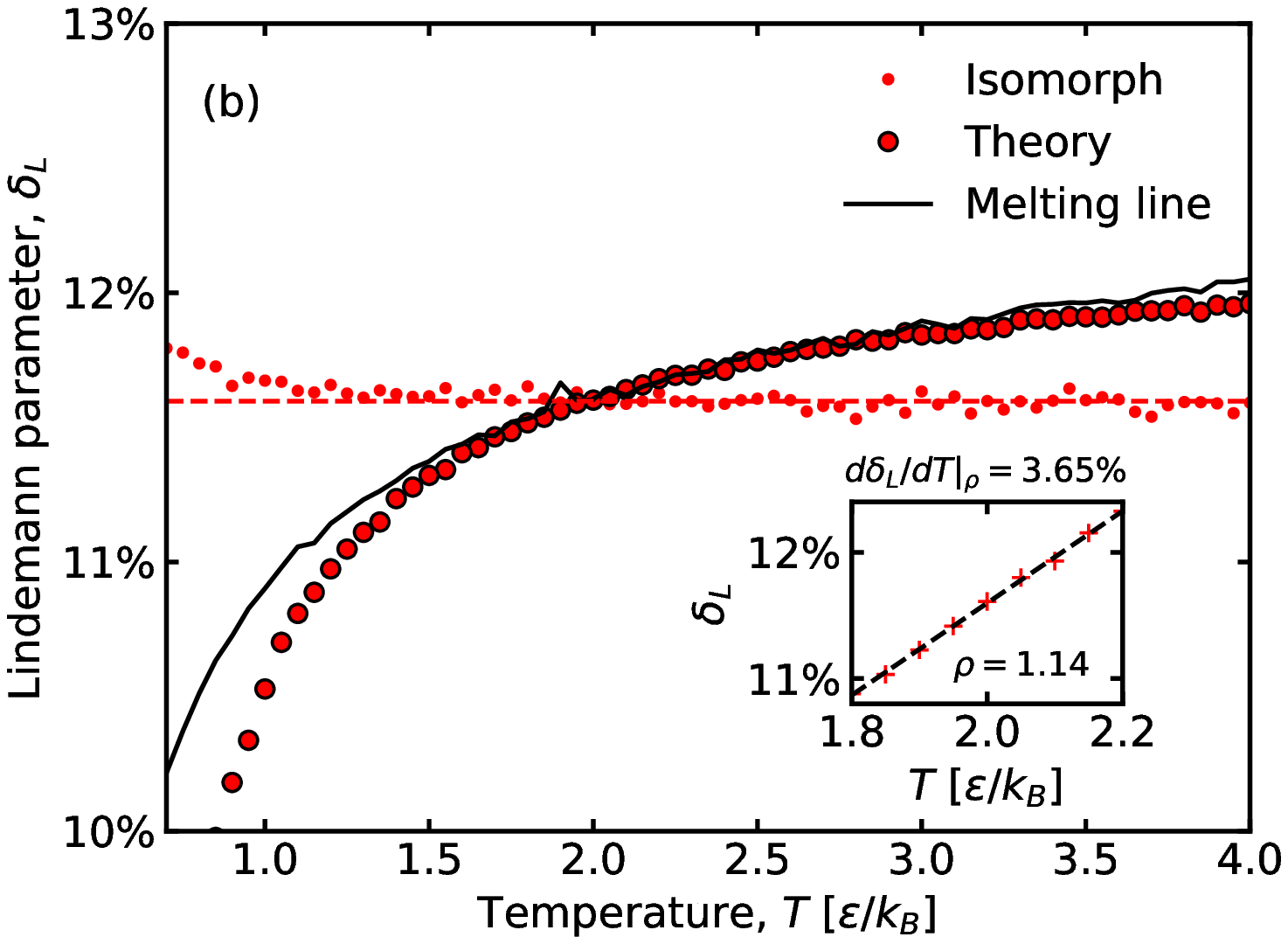}
		\end{center}
		\caption{\label{fig:melting_properties} (a) Predicting the entropy of fusion per particle $\Delta S_\textrm{m}/N$ and (b) the Lindemann parameter $\delta_L$ for the fcc solid from simulations at the $T_0=2\varepsilon /k_B$ reference state point. }
	\end{figure}
	
	For the Lennard-Jones system as an example, Ref.\ \cite{pedersen2016} showed how the freezing and melting lines, as well as the variation of several properties along these lines, could be calculated from simulations carried out at a single coexistence state point and by knowing the solid and liquid isomorphs through this state point.
	It is shown that the coexistent pressure as a function of temperature, $p_m(T)$, can be computed from a liquid- and a solid isomorph through a reference state point on the coexistence line with temperature $T_0$:
	\begin{equation}\label{eq:pm}
	p_m(T) = [C_1(T)+C_2(T)+TC_3]/C_4(T)
	\end{equation}
	where $C_1(T)$ is the difference between $U_s^I(T)-[T/T_0]U_s^I(T_0)$ and the analogous term for the liquid isomorph, $C_2(T)$ is the difference between $Nk_BT\ln(V_s^{(0)}/V_s^I(T))$ and the analogous term for the liquid isomorph, $C_3$ is the constant $p_0[V_l^{(0)}-V_s^{(0)}]/T_0$ and $C_4(T)=[V_l^I(T)-V_s^I(T)]$ is the volume difference between the two phases. Here the superscript ``I'' indicates values along the isomorphs, and ``0'' to values at the reference state point. On Fig.\ \ref{fig:melting_theory}(a) we test the melting theory for Argon using a liquid and a solid isomorph at the reference temperature $T_0=2\varepsilon/k_B$ (shown on Fig.\ \ref{fig:melting_theory}(b)). The theoretical prediction is good.
	
	Not only the shape in the pressure-temperature plane can be found, but also various properties along the freezing and melting lines -- again only using information from the reference state point.
	The density of the liquid at the freezing line $\rho_l$ for a given temperature $T$ is given by $\ln\rho_l=\ln\rho^I_l+(W_m-W^I)/(\partial W/\partial \ln\rho)_T$ where $W_m = p_m/\rho_l-k_BT$. The density of the solid $\rho_s$ is found by the analogous expression. The theoretical prediction is shown in Fig.\ \ref{fig:melting_theory}(b) as dots. The prediction is good though some deviation is notable at low temperatures. Similar deviations (but small) were also reported for the Lennard-Jones potential \cite{pedersen2016}. Some deviation from the theory is expected since the correlation coefficient is lower near the triple point.
	
	Figure \ref{fig:melting_properties}(a) shows the entropy of fusion $\Delta S_m$ per particle along the coexistence line (solid line). The theoretical prediction, shown as red dots, is remarkable. In comparison the hard-sphere picture predicts that the entropy of fusion is a constant.
	Figure \ref{fig:melting_properties}(b) shows the Lindemann parameter $\delta_L=[4/\rho]^{1/3}\sqrt{\Delta r^2/6}$ where $\Delta r^2=\langle|{\bf r}_i(0)-{\bf r}_i(t\rightarrow\infty)|^2\rangle$ is the root mean squared displacement of particles in the crystal at long times \cite{luo2005} ($\langle u^2\rangle= \Delta r^2/3$ in Reference \cite{luo2005}). The $+$'s on the insert show the Lindemann parameter along the $\rho=1.14$ isochore. The dashed line is a linear fit that yields $\partial \delta_L/\partial T|_{\rho}=0.0365k_B/\varepsilon$ needed in the theoretical prediction. The theoretical predictions are good for both $\Delta S_\textrm{fus}$ and $\delta_L$, however, they become less accurate at low temperatures. This is related to the bad prediction for the solid density near the triple point shown on Fig.\ \ref{fig:melting_theory}(b).
	
	\section{Conclusions}
	In summary, we have investigated the solid-liquid coexistence of argon in view of the hidden-scale invariance of the SAAP potential and shown that it can be used to give theoretical predictions of properties along the coexistence line. In the second paper of this series we extend the investigation to the noble elements Ne, Kr and Xe.
	
	\section{ACKNOWLEDGMENTS}
	The authors thanks Ian Bell, Søren Toxværd, Lorenzo Costigliola, Thomas B.\ Schrøder and Nicholas Bailey for their suggestions during the preparations of this manuscript and support by the VILLUM Foundation’s Matter grant (No. 16515).
	
	\section{SUPPLEMENTARY MATERIAL}
	See the supplementary material in Zenodo.org at \url{http://doi.org/10.5281/zenodo.3888373} for the raw data, and additional graphs.
	
	\section{Data Availability}
	The data that support the findings of this study are openly available in Zenodo.org at \url{http://doi.org/10.5281/zenodo.3888373}, and the supplementary material.
	
	\bibliography{references}
\end{document}